%% file: main.tex
\documentclass[useAMS,usenatbib,letterpaper]{mn2e}

\usepackage{aas_macros,amsmath,graphicx,lineno,longtable,amssymb,lastpage,mydefs,bm,lastpage}
\usepackage[version=3]{mhchem}
\title{What and Whence \obj: A Contact Binary from the Debris of a Young Planetary System?}

\author[Gaidos]{E. Gaidos$^{1}$\\
$^{1}$Department of Geology \& Geophysics, University of Hawaii at M\={a}noa, Honolulu, Hawaii 96822 USA\\
}

\begin{document}
\date{Submitted to MNRAS}
\pagerange{\pageref{firstpage}--\pageref{lastpage}} \pubyear{2017}

\maketitle

\label{firstpage}

\begin{abstract}
The first confirmed interstellar interloper in our Solar System, \obj{}, is likely to be a minor body ejected from another star, but its brief flyby and faintness made it difficult to study.  Two remarkable properties are its large (up to 2.5 mag) rotational variability and its motion relative to the Sun before encounter.  The former suggests an extremely elongated shape (aspect ratio $\ge$10) and the latter an origin from the protoplanetary disk of a young star in a nearby association.  Against expectations, it is also not comet-like.  \obj's variability can also be explained if it is a contact binary composed of near-equilibrium ellipsoidal components and heterogeneous surfaces, i.e. brighter, dust-mantled inner-facing hemispheres and darker, dust-free outer-facing poles.  Such shapes are a plausible outcome of radiation, tides and collisions in systems where planets are clearing planetesimal disks.  The probability that \obj{} has the same motion as a young ($\lesssim 100$~Myr) stellar association by coincidence is $<1$\%.  If it is young, its detection vs. more numerous, older counterparts could be explained as a selection effect due to darkening of surfaces by Galactic cosmic rays and loss of dust.  \obj's apparent lack of ices can be explained if ejected rocky planetesimals are characteristically smaller and thus far more numerous than their icy counterparts:  the Solar System may currently host several such objects captured by the combined gravity of Jupiter and the Sun. 

\end{abstract}


\input{intro}

\input{shape}

\input{origin}

\input{discussion}

\section*{Acknowledgements}

I thank Jeff Gillis-Davis, Paul Lucey, Sasha Krot, and Gary Huss for their advice and assistance, Wesley Fraser for providing a fresh perspective on the manuscript, and a second, anonymous reviewer for helpful suggestions.

\newpage
%\bibliographystyle{mn2e}
%\bibliography{references} 

\end{document}

%% file: intro.tex
\section{Introduction}
\label{sec.intro}

We have two windows on the formation and early evolution of planets; the study of meteorites and asteroids in our Solar System, some of which are fragments of primitive bodies that escaped incorporation into planets, and observations of circumstellar disks around young stars in which planets are, presumably, forming by aggregation of solids.  Through the former window, we can study relatively primitive protoplanetary material up close and determine the timing of some events;  however the processes by which these materials are altered and accreted into planets are not directly observed, nor are we certain this is an unbiased sample of protoplanetary material.  Through the latter window, we observe the formation of disks and the coagulation of dust grains, as well as the final products (exoplanets), but most intermediate steps are obscured because the largest objects, containing most of the mass, make the smallest contribution to emitting surface area; some steps could also occur too rapidly to be represented.  Data on the solid body population of young planetary systems is key to understanding their formation and the initial and environmental conditions that control outcomes.

 \obj{} (alias A1/2017~U1) is the first small body discovered with an unambiguously hyperbolic and hence unbound orbit \emph{before} entering the Solar System.  It was discovered in Pan-STARRS images after perihelion and soon after its closest approach to Earth, recovered in images taken earlier by the Catalina Sky Survey, and monitored for orbit and lightcurve determination by telescopes worldwide \citep{Meech2017}.  \obj{} lacked a coma and thus near-surface ices, but its reddish color is consistent with low-albedo asteroids and comets \citep{Meech2017,Jewitt2017,Ye2017,Bannister2017,Bolin2017,Fitzsimmons2017,Drahus2017}.  It is $\sim$200~m in size assuming a low albedo.  

Probably its most remarkable property besides its orbit is its 2-2.5 magnitudes of apparent photometric variability (uncorrected for phase effects), presumably due to rotation ($\approx$ 7-8~hr) of a non-spherical shape \citep{Meech2017,Jewitt2017,Knight2017,Bannister2017,Drahus2017}.  An ellipsoidal body (semi-axes $a$, $b$, $c$) with a uniform surface rotating around its shortest ($c$) axis typically produces a light curve that varies approximately as $a/b$; less if viewed nearly pole-on.  This has led to the deduction that the aspect ratio of \obj{} is at least 10:1 \citep{Meech2017}, although \citet{Fraser2017} points out the phase-angle effects could reduce this value.  No known body in the Solar System has such an extreme shape, and it requires that \obj{} has a small but non-zero cohesive strength. 

Among Solar System minor bodies with the largest photometric variability are contact binaries where, at certain viewing geometries, one component occults and/or eclipses (shadows) the other.  These eclipses produce characteristic cusp-like minima in the lightcurves like that seen in photometry of \obj{}.  Binaries, including contact binaries and ``dumb-bell'' shapes, occur in all minor body populations of the Solar System \citep{Walsh2015}.  A binary configuration, if confirmed, has implications for the size, internal properties and history of this object.

Another remarkable property of \obj{} is its pre-encounter velocity, which is within a few \kms{} of groups of nearby young ($\lesssim 100$~Myr) stars (Young Moving Groups or Stellar Associations), and not too different from the Local Standard of Rest.  This further solidifies \obj's extrasolar origin \citep{Mamajek2017} and raises the possibility that it is an ejected planetesimal from a young star system \citep{Gaidos2017}, and thus a representative of a population of planetary building-blocks not observationally accessible by other means.  (See \citet{Hansen2017} and \citet{Rafikov2018} for an alternative origin around a white dwarf).  Here, I examine the rotation, shape, surface, and origin of \obj{} in this context, and discuss the implications of its properties for its formation, age, and the parent population of ejected objects.

%% file: shape.tex
\section{Rotation and Shape}
\label{sec.shape}

To revisit the question of \obj's rotation and shape, photometric data of \obj{} were compiled from \citet{Meech2017}, \citet{Jewitt2017}, \citet{Knight2017}, and \cite{Bannister2017}.   Apparent magnitudes in \cite{Jewitt2017}, \citet{Knight2017}, and \citet{Bannister2017} were corrected for differences in solar illumination and distance changes relative to October 25.0 using the JPL Horizons calculator.  Magnitudes from those latter sources were converted to $g$-band using the colors estimated by those works; conversion (back) to the Sloan from the Cousins photometry reported by \citet{Jewitt2017} used the transformations of \citet{Jordi2006}.  

\begin{figure}
\centering
\includegraphics[width=\columnwidth]{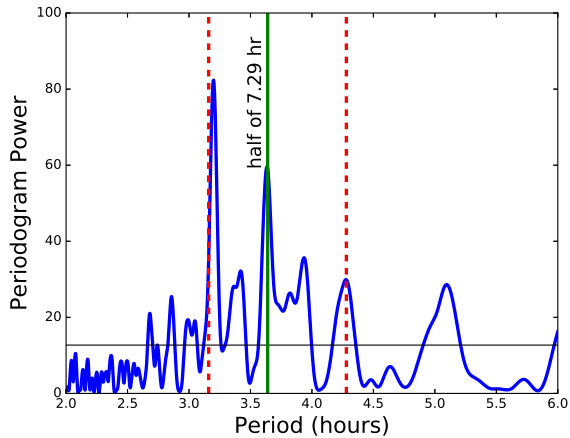}
   \caption{Lomb-Scargle periodogram of the combined photometry of \citet{Meech2017}, \citet{Jewitt2017}, \citet{Knight2017} and \citet{Bannister2017}.  The green line denotes the period close to that identified by \citet{Meech2017}.  The red dashed lines mark the alias signals produced by 24~hr sampling.}
\label{fig.period}
\end{figure}

The two highest peaks in the Lomb-Scargle periodogram \citep{Scargle1982} of the entire data set (Fig. \ref{fig.period}) are at 3.20 and 3.64 hr.  For rotation of a non-spherical shape, the most prominent signal is interpreted as one half the rotation period, in this case 6.40 and 7.29 hr.  The latter peak is very close to the period identified by \citet{Meech2017}.  However, sampling at an interval $S$ can produce alias signals at $1/(1/P \pm 1/S)$ and for ground-based observations with $S=24$~hr, at least one of the peaks could be an alias (Fig. \ref{fig.period}).  

\begin{figure}
\centering
   \includegraphics[width=\columnwidth]{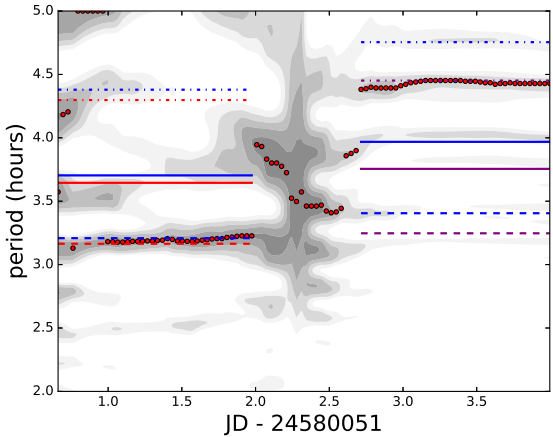}
   \caption{Periodogram power vs. time where the moving window is adjusted to always include 100 data points and time is calculated as the mean time of points in the window.  Red points mark period of maximum power.  The red line is the half-period corresponding to a rotation of 7.29~hr.  Blue and purple solid lines are the rotation half-periods proposed by \citet{Fraser2017} and this work, respectively.  Dashed lines with corresponding colors are the expected alias signals produced by 24-hr sampling.}
\label{fig.scan}
\end{figure}

Moreover, the phased data reveal inconsistencies between the periodicity at earlier and later times.  Fig \ref{fig.scan} shows a running periodogram computed with a moving window having a width that is adjusted to always include 100 data points.  The time assigned to each periodogram is the mean time of the data in the window.   A consistent pattern before relative $\Delta$JD $=2$ (``early") is replaced by a different pattern after $\Delta$JD $=2.7$ (``late").  At intermediate times, the data cannot be explained by periodic signals, presumably because the window is sampling different periods in both early and late intervals.  In neither of those intervals is the 7.29-hr signal dominant, a result of limited and biased sampling of the signal.  In the early interval the alias at 6.3~hr dominates.  In the latter interval, the dominant 4.4~hr signal is \emph{not} an alias of 7.29~hr and in fact no single period and its aliases can reproduce the full pattern of signals. 

Based on such discrepancies, \citet{Fraser2017} proposed that \obj{} is a non-principal axis (NPA) rotator and ``tumbling".  They found that two periods of 7.4 and 7.9~hr (plotted as the solid blue lines in Fig. \ref{fig.scan}) can adequately, but perhaps not uniquely explain the data.  These values can explain the short-period alias in the early interval, but not the 8.8~hr signal in the latter interval, except, perhaps if one invokes a 48~hr sampling alias.  A different choice of 7.29 and 7.51~hr periods (solid purple lines) better explains the 6.3 and 8.8~hr signals, but not the (weaker) 8~hr signal.   The 7.51~hr period is close to the 7.55~hr found by \citet{Drahus2017} in the late interval.

If NPA rotation occurs, changes in rotation period constrain the ratio of principal axes.  Because angular momentum is conserved, the ratio of periods $P_1/P_2$ is proportional to the ratio of principal moments of inertia $I_1/I_2$. In the case of an ellipsoidal geometry, $P_1/P_2 = (a^2 + b^2)/(a^2 + c^2)$, where rotation alternates between the $b$ and $c$ axes.  When $a \gg b,c$, as has been proposed for \obj{}, $P_1/P_2 \approx 1 + (b^2-c^2)/a^2$.  For such extremely elongated objects the long axis dominates the moment of inertia and the difference in periods is small.  If $a/b = 10$ and $b \approx c$, the difference is $\ll1$\%, i.e. much less than 4 minutes.  The difference in period implied by the observations is much larger. e.g. about 30~min \citep{Fraser2017},  16~min  \citep{Drahus2017}, or 13~min (this work).  A difference of at least 3\% indicates $a/b < 5$.  Real asteroids are of course not perfectly ellipsoidal, but the scaling relationships still hold, the long axis will still force the difference in moments of inertia to be small, and deviations from the ellipsoid case will only modulate that difference in a minor way.  For example, the difference between the individual principal moments of inertia of a cuboid is only 8\% larger than that of an ellipsoid with the same volume and axis ratios.

\begin{figure}
\centering
   \includegraphics[width=\columnwidth]{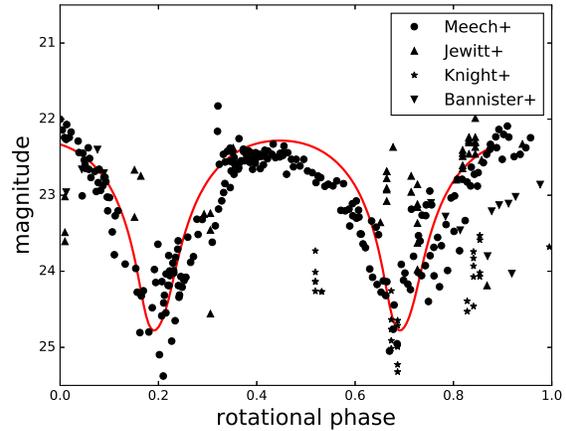}
   \caption{Lightcurve of \obj{} produced by phasing published photometry to a 7.29 hr period.  The solid red curve is a model of a contact binary composed of equal-mass equilibrium prolate ellipsoids with non-uniform surfaces (Hapke-scattering except for dark Lambertian-scattering ``caps" on the outer poles extending 57~deg to the equators, see text.)  Arbitrary offsets of rotational phase and magnitude have been applied to the model.}
\label{fig.photometry}
\end{figure}

Photometry of \obj{} phased to 7.29 hr are plotted in Fig. \ref{fig.photometry}), showing the apparent amplitude $\Delta m \approx 2.5$ (uncorrected for phase effects) which inspired aspect ratio estimates of $a/b \approx 10$.  However,  the amplitudes of asteroid lightcurves depends on phase angle in a manner that is a function of geometry and surface scattering properties \citep[e.g.,]{Gutierrez2006}, and this effect must be included when estimating its shape \citep{Fraser2017}.  To ascertain this, I calculated lightcurves using a finite-element model of a rotating ellipsoid having $a>b=c$ and a uniform surface following either a Lambertian or Hapke scattering law including macroscopic roughness \citep{Hapke1993}, and adopting the Earth-Sun-\obj{} viewing geometry of 27 October 2017 (phase angle = 22 deg).  The lightcurve amplitude $\Delta m$ depends on the unknown rotational pole vector of \obj{}, but the maximum possible amplitude for a given $a/b$ can be found by varying obliquity ($\delta$) and longitude of the rotation axis with respect to the orbital apse line ($\omega$).  

\begin{figure}
\centering
   \includegraphics[width=\columnwidth]{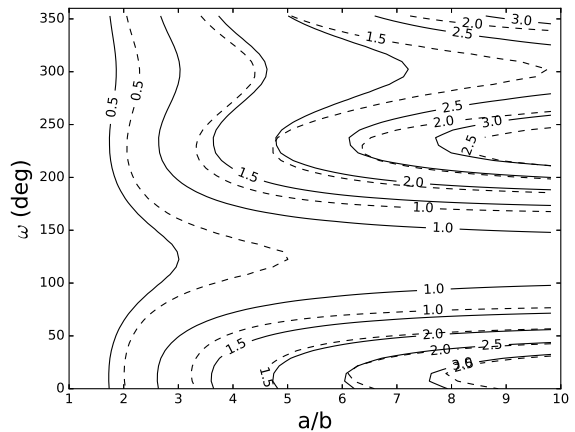}
   \caption{
   Rotational lightcurve amplitude (magnitudes) of a single prolate ellipsoid vs aspect ratio $a/b$ ($b=c$) and longitude of the rotation axis $\omega$ with respect to the orbit apse, for the case of a Lambertian scattering surface (solid contours) and Hapke scattering with parameters of the average C-type asteroid (dashed contours).  The rotational obliquity $\delta$ is set to 30~deg.: varying $\delta$ changes the dependence on $\omega$ but largely not on $a/b$.}
\label{fig.deltamag1}
\end{figure}

Figure \ref{fig.deltamag1} plots $\Delta m$ as a function of $a/b$ and $\omega$ for $\delta = 30~$deg.  Solid contours are for a Lambertian surface and dashed contours are for Hapke scattering with parameter values characteristic of C-type asteroids \citep{Li2015}.  The maximum $\Delta m$ occurs when the object's aspect angle (angle between the rotation axis and line of sight) is near 90~deg and is fairly insensitive to $\delta$, although the range of satisfactory values of $\omega$ increases with decreasing $\delta$. With a Lambertian surface, $\Delta m = 2.5$ can be produced for $a/b$ as small as 6, but for that of a C-type asteroid, the minimum $a/b = 8$.  (Average parameters for S- and V-type asteroids give similar results).  \citet{Drahus2017} find a smaller lower limit of $a/b = 4.63$. 

Large lightcurve amplitudes as well as the cusp-like minima seen in Fig. \ref{fig.photometry} are characteristic of binary asteroids where one component eclipses the other twice a rotation/orbit.  The possibility that \obj{} is a contact binary was considered and rejected by \citet{Meech2017} but revisited by \citet{Bannister2017}.  Contact binaries have representatives among all small body populations in the Solar System \citep[e.g.,][]{Hudson1997,Margot2002,Marchis2006,Marchis2014,Waszczak2015}.  Mechanisms that can produce such objects include spin-up and rotational disruption as a result of the Yarkovsky-O'Keefe-Radzievskii-Paddack (YORP) effect \citep{Pravec2007}, tidal disruption during a close encounter with a planet \citep{Mordidelli2006} or a collision \citep{Paolicchi2002}.  The requirements of any of these mechanisms (proximity to a star, encounter with a planet, or collision with another body) are consistent with the scenario where \obj{} formed close to another star and was ejected by a planet (see Sec. \ref{sec.discussion}).

A body formed as a result of disruption is expected to have very low intrinsic strength and its shape could closely follow an equipotential surface.  This configuration can be approximated by a pair of Roche ellipsoids with axis ratios that depend on the mass-ratio of the components and the ratio of centripetal to gravitational forces $\Omega^2/(\pi G\rho)$, where $\Omega = 2\pi/P$, $G$ is the gravitational constant, and $\rho$ is the mean density \citep{Leone1984}.  The maximum axis ratio occurs at the Roche stability limit ($a/b \approx 1.6$) where, for equal-mass components $\rho \approx 2.1$~g~cm$^{-3}$ (P/7.3~hr)$^{-2}$.  Many asteroids have this density or higher \citep{Carry2012}.  The axis ratio of each component of such a binary is about 1:0.62:0.56 \citep{Leone1984}.  A contact binary can also tumble:  For such an object the difference in periods would be 15 minutes.

\begin{figure}
\centering
   \includegraphics[width=\columnwidth]{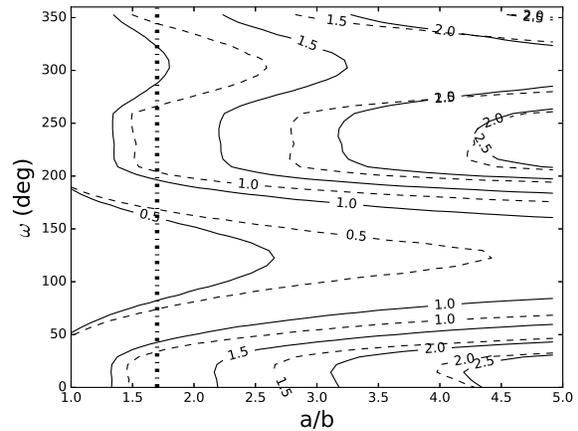}
   \caption{Rotational lightcurve amplitude (magnitudes) of a contact binary consisting of two identical prolate ellipsoids vs. individual aspect ratio $a/b$ ($b=c$) and longitude of the rotation axis $\omega$ with respect to the orbit apse, for the case of a Lambertian scattering surface (solid contours) and Hapke scattering with parameters of the average C-type asteroid (dashed contours).  The rotational obliquity $\delta$ is set to 30~deg.: varying $\delta$ changes the dependence on $\omega$ but largely not on $a/b$.  The vertical line is the expected axis ratio of a Roche ellipsoid at the stability limit.}
\label{fig.deltamag2}
\end{figure}

Figure \ref{fig.deltamag2} is the same as Fig. \ref{fig.deltamag1}, but \obj{} is modeled as a contact binary consisting of identical prolate ellipsoids aligned along the connecting axis.  Bodies at the stability limit $a/b\approx1.7$ (vertical dashed-dot line) cannot reproduce the observed variability of \obj{}.  Larger aspect ratios and greater variability are possible if the bodies have finite strength, i.e. a minimum $\sim 1$~Pa \citep{Meech2017,Bolin2017}.  For a Lambertian-scattering surface (solid contours in Fig. \ref{fig.deltamag2}), a minimum ratio of $a/b = 4.2$ is required to explain $\Delta m = 2.5$ and thus an overall ratio of $>8.4$).  Hapke scattering characteristic of a C-type asteroid \citep[dashed contours,][]{Li2015} requires still larger ($\approx10$) axis ratios.   These values could conflict with the observation of significantly different periods during tumbling. 

\begin{figure}
\centering
   \includegraphics[width=\columnwidth]{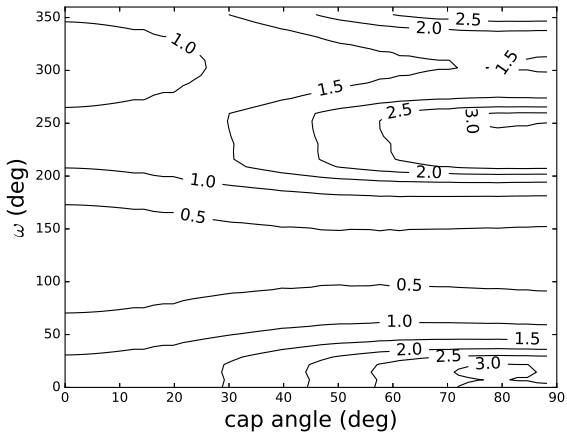}
   \caption{Rotational lightcurve amplitude (magnitudes) of a contact binary consisting of two Roche ellipsoids $a/b=1.7$ ($b=c$) with end-caps of C-type scattering surfaces and brighter E-type asteroid-like material over the remainder.  The axes are the angle subtended by the caps relative to the outer poles of the ellipsoids and the longitude $\omega$ of the rotation axis with respect to the orbit apse, a proxy for the aspect angle.  The obliquity is fixed at $\delta = 30$~deg.}
\label{fig.deltamag3}
\end{figure}

More modest axis ratios combined with surface reflectance variation can explain the lightcurve of \obj.  In particular, if the outer poles of the bodies are much darker than the inward-facing surfaces, the axis ratios can be consistent with an equilibrium ellipsoid.  Figure \ref{fig.deltamag3} plots $\Delta m$ vs. the angular size of the dark end-caps (as measured from the poles).  The surface of the caps is modeled as a Lambertian scatterer with albedo of 0.03 and the remaining surface as a Hapke scatterer with the parameters of a relatively bright E-type asteroid \citep[geometric albedo = 0.34]{Li2015}.  Such extreme (factor of 10) albedo variation has not been observed for any Solar System asteroid, but is comparable to that of Pluto and the Saturnian satellite Iapetus.  A dark region extending 60~deg from the outer poles of the surface will generate the observed variability (Fig. \ref{fig.deltamag3} and red curve in Fig. \ref{fig.photometry}).  Differences between the observations from the model are most likely due to the departures of the object's shape from a pair of ellipsoids and more complex variations in albedo than considered here.   

One reason the albedo could be non-uniform is a partial mantle of brighter dust covering a darker surface.  The latter is absent where the net surface acceleration (gravity minus centripetal force) is weakest, i.e. at the outward-facing poles of the components \footnote{Although the shape of a strengthless body can relax to an equipotential surface, the net acceleration at the surface -- the vertical gradient of the potential -- will vary.} Although the reflectance spectra of the two different surfaces may differ markedly, this might not be apparent in a low signal-to-noise spectrum since the signal from the brighter surface component dominates.  

%% file: origin.tex
\section{Origin of \obj}
\label{sec.origin}

The radiant point of \obj{} (near 15h48m +34d01') is only 8 deg. from the Solar apex, reflecting its low ``peculiar" velocity with respect to the Local Standard of Rest (LSR).  Molecular clouds and newly formed stars largely move with the LSR, but as stars age their peculiar velocities increase as roughly $t^{1/2}$ due to scattering by other clouds and asymmetries in the Galactic potential such as spiral arms \citep[e.g.]{Aumer2016}; this is the basis of the kinematic method of estimating stellar ages.  Also, coeval groups of stars spawned by the same cloud initially share a common space motion ($UVW$) this can be used to identify members of these associations for up to hundreds of Myr, long after they have spatially dispersed \citep[e.g.,][]{Gagne2018}.

The pre-encounter space motion of \obj{} ($UVW = -11.440 \pm 0.009$,$-22.377\pm0.009$, $-7.743\pm0.010$~\kms) and its peculiar motion is 9~\kms{} with respect to the LSR \citep{Coskunoglu2011}, much less than the 50 \kms{} velocity dispersion of nearby stars in the Galactic disk \citep{Anguiano2017}.  Most of this 9~\kms{} is in the direction of Galactic rotation ($V$) where the uncertainty in LSR motion is largest.\footnote{The Sun was likely \obj's first stellar encounter since ejection since the probability of approaching within 100~AU of any star in 5~Gyr is $\lesssim 1$\%.}  Based on this, \citet{Gaidos2017} and \citet{Feng2018} suggested that the formation (or at least ejection) age of \obj{} is young compared to stars in the Solar neighborhood\footnote{Statistically, \obj{} would have an equally high peculiar motion whether it was recently ejected from an old star or ejected long ago from a young star that is now equally old.}  Moreover, the space motion of \obj{} is within 1-2 \kms{} of that of the 40-50~Myr-old Carina and Columba stellar associations, leading \citet{Gaidos2017} to suggest an origin around a star in a member of these associations and ejection sometime after 40-50~Myr ago.  An object moving at 1-2~\kms{} would cover the distance to these clusters ($\approx$50~pc) in under 50 Myr.   

However, the correlation between the kinematics of stars (and ejected planetesimals) and age is only a \emph{statistical}  relation and, like old stars, \obj{} could have a low peculiar motion close to that of a young stellar association by coincidence.  A frequentist calculation of the probability of this null hypothesis is:
\begin{equation}
p = \frac{d_u d_v d_w }{\sigma_u \sigma_v \sigma_w}\sum_i \exp\left[-\Delta \vec{v_i} \bm{\mathit{D}} \Delta \vec{v_i} / 2\right],
\end{equation}
where $\bm{\mathit{D}}$ is the inverse of the velocity dispersion matrix of the cluster, $\Delta \vec{v_i}$ is the relative velocity of the $i$th association with respect to the local standard of rest \citep{Coskunoglu2011}, $d$ are the principal components of $\bm{\mathit{D}}$, and $\sigma$ the velocity dispersion components of the thin disk \citep{Anguiano2017}.  (Errors in the space motion of \obj{} are 10~\mps{} and can be neglected.)   The associations included in the summation are TW Hydra, $\beta$ Pictoris, Tucana-Horologium, Columba, Carina, Argus, and AB Doradus \cite[from][]{Gagne2014}, $\sigma_{32}$ Orionis \citep{Bell2017}, $\epsilon$ and $\eta$ Chamaeoleontis \citep{Mamajek1999,Murphy2013}, Octans \citep{Torres2008}, $\alpha$ Persei \citep{Makarov2006}, the $\rho$ Ophiucus and Upper Scorpius star-forming regions \citep{Dahm2012,Cook2017} and the Pleiades \citep{Galli2017}.  The total probability is 0.4\%.  Although this is a lower limit, since there may be undiscovered nearby associations, the probability that the young association-like motion of \obj{} is a coincidence is very likely $<1$\%.  The space motion of the source association will not exactly equal that of \obj{} because of stellar velocity dispersion (typically $\sim$1~\kms{}) and a finite ejection velocity; either that latter is also small, or the association-like motion of \obj{} is a low-probability coincidence.  

\begin{figure}
\centering
   \includegraphics[width=\columnwidth]{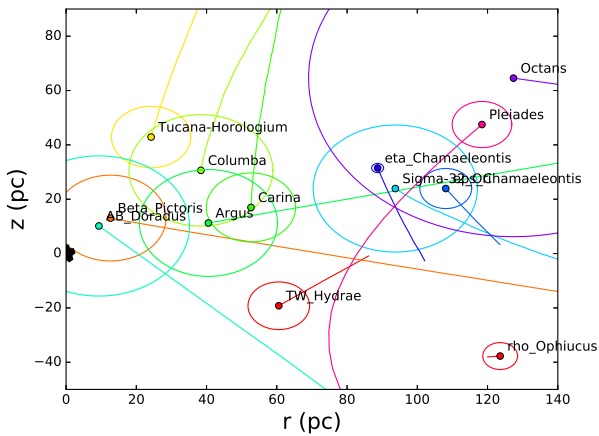}
   \caption{Plot of the relative motion of young stellar associations and clusters with respect to \obj{} (black star at origin) in $r \equiv \sqrt{x^2 + y^2}$ vs $z$ coordinates.  Circular points indicate the current centers of the clusters/association, the ellipses indicate the approximate current size, and the trails indicate the motion back to the estimated time of association/cluster formation.}
\label{fig.trackback}
\end{figure}

A more rigorous test of common origin would be to dynamically track the \emph{spatial} locations of both \obj{} and an association back in time to some epoch to where they coincide.  This is difficult because associations are often dispersed over tens of pc, their centers are not well determined, and the member stars could have had a finite velocity dispersion at the epoch of ejection.  Moreover, errors in position due to errors in velocity and perturbations by field stars \emph{accumulate} as $t^{3/2}$ to pc-scale errors in $\sim100$~Myr \citep{Zhang2018}.  
Figure \ref{fig.trackback} shows the results of track-back simulations for the different moving groups using the gravitational potential of \citep{Irrgang2013}.  The positions of the associations with respect to \obj{} in the rotating frame of the LSR are plotted as $r \equiv \sqrt{x^2 + y^2}$ and $z$, where $x,y,z$ are towards the Galactic center, direction of Galactic rotation, North Galactic Pole, respectively.  The tracks end at the estimated formation time of each association/cluster.  A potential origin would have a track that passed through or near the origin at the epoch of ejection.  The current physical sizes of the associations are indicated by the ellipses in Fig. \ref{fig.trackback}, but this does not include the spatial extent corresponding to the accumulation of velocity uncertainty with time. 

No unambiguous candidate for the birthplace of \obj{} emerges from this analysis.  \obj{} is still \emph{approaching}, not moving away from, the Carina, Columba, and Tucana-Horologium clusters all of which have similar space motions.  \obj{} -- and the Sun -- are currently encountering the margins of the AB Doradus and $\beta$~Pictoris associations but the velocity differences are 9 and 7~\kms{}, respectively.  For the orbits of Carina and \obj{} to meet in the past the association's $v$ must be changed by 4 \kms{} and its $w$ by 3.2~\kms{}, something excluded by measurement precision.  Alternatively, a hypothetical association with the space motion of Carina but at a different current location could be the source.  A range of $xyz$ values are possible given the proportionally large uncertainty in the relative velocity between \obj{} and Carina and the dependence on the epoch of ejection, but one set is (29,6,82)~pc, or 87 pc in the direction of $\alpha = 212$~deg., $\delta = +19$~deg. (near the star Arcturus).  The Coma Berenices association is near this location but has a very different $UVW$ \citep{Gagne2014}.

%% file: discussion.tex
\section{Discussion}
\label{sec.discussion}

\obj{} almost certainly formed around another star and its detection implies a space density of similar objects corresponding to an ejected mass of a few \mearth{} per star \citep{Gaidos2017,Trilling2017,Raymond2017}. The circumstances and timing of \obj's formation and evolution in its home system, as well as the launch onto its interstellar voyage are not known, but \obj's shape, composition, and motion provide tantalizing hints.

Planetesimals are more likely to be ejected rather than accreted by a planet with an escape speed exceeding its orbital speed and thus more massive planets on wider orbits are likely to be responsible \citep{Gaidos2017,Raymond2017}.  Also, planets that are migrating or excited onto highly eccentric orbits can eject larger numbers of planetesimals.  The formation of a giant planet by rapid capture of a gas envelope, migration through (or with) the disk, and scattering by mutual gravitational perturbations are therefore potential major episodes of planetesimal ejection.  Gas giant formation and migration occurs in $\lesssim 10$~Myr (the typical disk lifetime) while orbital instability in two (giant) planet systems occur on a timescale that is a super-exponentially increasing function of separation from the nearest mean-motion resonance \citep[][the existence of additional planets complicates this picture]{Chatterjee2008}.  Planetesimal ejection, averaged over many systems, is thus expected to peak within 10~Myr and decrease with time.  This might be reflected in the decline in incidence of debris disks with age \citep{Carpenter2009,Sierchio2014}, if the dust is the product of collisions of planetesimals exited onto crossing orbits by giant planets.

Binaries, including contact binaries are one outcome of disruption of a minor body by Yarkovsky-O'Keefe-Radzievskii-Paddack (YORP) spin-up \citep{Pravec2007}, tidal disruption during a close encounter with a planet \citep{Bottke1996,Walsh2006} or a collision with another minor body \citep{Paolicchi2002}.  Radiation-driven (YORP) spin-up to the break-up rate requires at least $\sim$1~Myr of proximity to the host star, depending on mean distance and luminosity \citep{Jacobson2014}.  The mean collision time among planetesimals in a disk is proportional to the characteristic size and inversely proportional to the surface density of solids, but for a minimum-mass nebula of 1~km bodies at 5~AU it is comparable to the ejection time ($\sim 10^5$~yr) by a Jupiter-mass planet \citep{Tremaine1993}.  Disruption by tides is expected to be comparatively rare:  \citet{Raymond2018} found that only 0.1-1\% of ejected planetesimals in their simulations had previously passed within two Roche radii of a planet.  However, tidal disruption of large ($\sim$100~km) bodies would produce many smaller objects that could dominate the ejecta population (see below).

In planetary systems with solar-like ratios of heavy elements, the amount of mass in condensible ices is about twice that in ``rocks" \citep{Lodders2003}.  Since icy planetesimals form farther from a host star they are more likely to be ejected than their rocky inner counterparts, hence the expectation that most interstellar interlopers are icy \citep{Raymond2017}.  \obj's surprising lack of detectable cometary activity despite a perihelion of 0.25~AU indicates lack of ices within 1~m of its surface, the predicted penetration depth of the thermal wave.  \citet{Jewitt2017} and \citet{Fitzsimmons2017} proposed that Galactic cosmic rays depleted ices from this zone.  But this mechanism is contradicted by long-period comets from the Oort Cloud, which experienced interstellar conditions at $>10,000$~AU for 4.5~Gyr but retain surface ices.  Moreover, if \obj{} is composed of strengthless material filling critical Roche ellipsoids then its density is $\approx$2~g~cm$^{-3}$, also disfavoring significant ice content.  

An ice-free \obj{} could be explained if ejected rocky planetesimals are more numerous, but are typically much smaller.  The size distribution of Oort cloud comets flattens markedly below 2~km and is negligible below 500~m \citep{Fernandez2012}.  If the characteristic sizes for ejected rocky and icy planetesimals are 100~m (\obj{}-like) and 1~km (comet-like), respectively, then the relative numbers would exceed 100:1.  Differences in size distribution with ice content could arise from variation in the lifetime of disk gas, which promotes (re-)aggregation of smaller bodies, or more efficient collisional fragmentation of planetesimals collisions closer to the star \citep{Kenyon2012b}.  \citet{Raymond2018} link tidal disruption by a planet to the lack of ices in \obj{} by the desiccation that can occur if multiple star-approaching orbits follow tidal disruption but precede ejection.  They propose that the rarity of such events among ejected planetesimals (0.1-1\% in their simulations) is balanced by the shear number of fragments produced. 

The age of \obj{} is not known: a pre-encounter velocity that is close to those of some young stellar associations with a false alarm probability $<1$\% hints at a common origin and age of $\lesssim 100$~Myr \citep{Gaidos2017,Feng2018} but track-back simulations are limited by the precision of available stellar space motions, the completeness of catalogs of nearby young stars, and the rapid accumulation of positional uncertainty with reversed time \citep{Zhang2018}.  The \emph{Gaia} mission will provide a far more complete and accurate inventory of nearby young associations and their dynamics \citep{Moraux2016}, and perhaps resolution of this question.  

A young age would create a conundrum:  The average age of stars in the Galactic disk is a few Gyr and if each of these also produced planetesimals in roughly equal numbers, the prior probability that the first object we discover is much younger is very low.  This discrepancy would have to be explained by strong (factor of $\sim 100$) selection against older objects.  Tidal disruption of planetesimals by a close encounter with a star is $\sim10^{-9}$ in 5 Gyr and disruptive spin-up by the YORP effect is inoperable beyond $\sim 100$~AU for 100-m objects like \obj{} \citep{Jacobson2014}.  Mechanical torques could be important if the original object is sufficiently irregular, but only on a timescale of several 100 Myr \citep{Hoang2018}.  Gravitational focusing provides a modest bias against older objects moving at higher speeds due to scattering by stars and molecular clouds.  This bias is a factor $1 + 2(v_{\rm orb}/v_{\infty})^2$, where $v_{\rm orb}$ is the orbital velocity at 1~AU and $v_{\infty}$ is the planetesimal approach velocity.  This factor was calculated for nearby young associations and clusters  assuming that the space motions of young planetesimals were distributed like that of the stars in each parent moving group \citep{Torres2008,Gagne2014}, and the motions of older objects were distributed according to the velocity ellipsoid of the the Galactic thin disk \citet{Anguiano2017}.  Enhancements by at most 2-3 are predicted (Table \ref{tab.bias}), well short of what is required.

\begin{table}
\begin{center}\label{tab.bias}
\caption{Dynamical Bias Ratios} 
\begin{tabular}{lc}
Stellar Group & Bias at 1~AU$^{a}$\\
\hline
Upper Scorpius & 2.8\\
$\rho$~Ophiucus & 2.3\\
Octans & 2.2\\
Beta Pictoris Moving Group & 1.7\\
TW Hydra Association & 1.5\\
Tucana-Horologium & 1.5\\
$\sigma_{32}$~Ori & 1.4\\
Columba & 1.3 \\
Argus & 1.3\\
Carina & 1.3\\
$\epsilon$~Chamaeleontis & 1.3\\
$\eta$~Chamaeleontis & 1.2\\
$\alpha$~Persei & 1.0\\
AB Doradus & 0.9\\
Pleiades & 0.9\\
\hline
\end{tabular}
\end{center}
$^a$relative to the velocity distribution of the nearby Galactic disk.
\end{table}

The surfaces of Main Belt asteroids become redder and darker from radiation and micrometeorite exposure (``space weathering") in $10^4-10^6$~yr \citep{Brunetto2015}, and an analogous process acting on interstellar objects could render them darker and undetectable as their surfaces age.\footnote{\obj{} was only observable when it was close to both Earth and the Sun and would not have been detected had it been significantly fainter. For example, if it were entirely covered by a dark material and lacked the brighter midriff that could explain the rotational variability, its \emph{peak} brightness would be 2.3 magnitudes fainter.}  The interstellar flux of protons in the energy range (up to a few keV) thought to be partly responsible space weathering is $\lesssim 10^-3$ of the flux at 2.5~AU \citep{Brunetto2015}, but the flux of higher energy Galactic cosmic rays is many orders of magnitudes greater due to the absence of a Solar magnetic field \citep{Cooper2003}.  Darkening could occur by loss of any brighter, scattering dust mantle, e.g. by electrostatic charging and levitation \citep{Mendis1981,Lee1996}.   This dust layer would not be replenished by micrometeorite impacts since the mass flux of impacting interstellar dust particles is at least 2-3 orders below that estimated in the inner Solar System \citep{Altobelli2016,Bennett2013}.  The corollary of this explanation is that many more objects with older, darker surfaces enter the Solar System but remain undetected. 

Drawing conclusion from a single object in a population is dangerous, especially because selection effects like those described above may influence a detection.  The addition of a second telescope to the Pan-STARRS survey \citep{Burgett2012,Engelhardt2017} and the advent of the Large Synoptic Survey Telescope in 2021 \citep{MoroMartin2009,Cook2016} should provide meaningful constraints on the size, shape, and velocity distribution of interstellar planetesimals and, perhaps, insight into planetesimal growth and planet formation from a new perspective.  If there is a strong detection bias towards young objects, many future detections will have radiants near the Solar apex.

A close-up investigation of an interstellar interloper like \obj{} is clearly desirable, but a space mission to rendezvous (or collide) with one would be challenged by the high $\Delta v$ required and the limited warning time.  Any interstellar objects captured onto a bound orbit around the Sun by Jupiter \citep[no other planet offers a comparable capture cross-section:][]{Valtonen1982} would be a more accessible target of exploration.  Based on a number density for extrasolar comets of $\sim 10^{13}$~pc$^{-3}$, \citet{Torbett1986} estimated one capture per 60~Myr, much longer than the ejection by Jupiter \citep[450 kyr,]{Levison1994}.  If the actual number density of \obj{}-like objects is higher, the steady state number could be a few.  

Alternatively, smaller interstellar objects could be numerous enough to occasionally strike Earth as meteorites.  In the case of the -7/2 power-law distribution produced by a collisional cascade \citep{OBrien2003}, for every \obj{}-size object that passes within 1~AU of the Sun, a sister 30-cm object collides with Earth.  About 1000 objects of such size impact every year \citep{Brown2002} and since there are about 1200 meteorite falls in world-wide collections \citep{Krot2014}, this statistic, subject to the assumption of a steep size distribution raises the possibility that one may be interstellar.  No extrasolar meteorites have been identified, but two indicators of extrasolar origin are a primitive chondrite-like composition combined with a non-solar Pb-Pb radiometric age, and an oxygen isotopic composition that differs markedly from that of the Solar System \citep{Clayton1993} due to Galactic chemical evolution \citep{Gaidos2009,Young2011}.